\documentclass{conm-p-l} 

\newtheorem{theorem}{Theorem}[section]
\newtheorem{lemma}[theorem]{Lemma}

\theoremstyle{definition}

\theoremstyle{remark}
\newtheorem{remark}[theorem]{Remark}

\numberwithin{equation}{section}

\newcommand{\Phh}{\Phi}
\newcommand{\opS}{{\mathbb S}}

\newcommand{\nonu}{\nonumber \\} 
\newcommand{\cE}{{\mathcal E}}
\newcommand{\cC}{{\mathcal C}}

\newcommand{\ii}{{\rm i}}
\newcommand{\ee}{{\rm e}}
\newcommand{\dd}{{\rm d}}
\newcommand{\mbf}[1]{\mbox{\boldmath ${#1}$}}

\newcommand{\vx}{{\bf x}}
\newcommand{\vy}{{\bf y}}
\newcommand{\vz}{{\bf z}}
\newcommand{\vn}{{\bf n}}
\newcommand{\vm}{{\bf m}}
\newcommand{\ve}{{\bf e}}
\newcommand{\vE}{{\bf E}}
\newcommand{\vmu}{\hat{\mbf \mu}}
\newcommand{\vzero}{{\mbf 0}}

\newcommand{\eps}{\varepsilon}
\newcommand{\half}{\mbox{$\frac{1}{2}$}}

\newcommand{\R}{{\mathbb R}}
\newcommand{\C}{{\mathbb C}}
\newcommand{\Z}{{\mathbb Z}}
\newcommand{\N}{{\mathbb N}}

\renewcommand{\equiv}{:=\,}

\begin{document}

\title[Explicit formulas for Jack polynomials]{A method to derive
explicit formulas for an elliptic generalization of the Jack
polynomials}

\author{Edwin Langmann}

\address{Mathematical Physics, KTH Physics, AlbaNova, SE-106 91
Stockholm, Sweden}

\email{langmann@theophys.kth.se}

\thanks{Supported by the Swedish Science Research Council~(VR), the
G\"oran Gustafsson Foundation, the {\it ``Knut och Alice Wallenbergs
Stiftelse''}, and the European grant ``ENIGMA''}

\copyrightinfo{2004}
    {American Mathematical Society}
       
\date{\today}


\begin{abstract}
We review a method providing explicit formulas for the Jack
polynomials. Our method is based on the relation of the Jack
polynomials to the eigenfunctions of a well-known exactly solvable
quantum many-body system of Calogero-Sutherland type.  We also sketch
a generalization of our method allowing to find the exact solution of
the elliptic generalization of the Calogero-Sutherland model. We
present the resulting explicit formulas for certain symmetric
functions generalizing the Jack polynomials to the elliptic case.
\end{abstract}

\maketitle

\section{Introduction}
\label{sec1}
In this paper we explain a method which yields explicit formulas for
the Jack polynomials; see \cite{St,McD}.  We found this method by
studying integrable quantum many-body systems of Calogero-Sutherland
type \cite{C,Su}, and our discussion will be therefore from the
quantum integrable systems' point of view.  However, we made an effort
to make this paper also useful to readers interested in symmetric
polynomials and not so much in the physics interpretation of our
results. To explain our notation (which is admittedly less elegant
than the one used in \cite{McD,St} but closer to the one used in
physics) we first give a definition of the Jack polynomials. (The
equivalence of our definition and the one in \cite{St} follows from
Theorem 3.1 in \cite{St}.)  We then recall the relation of the Jack
polynomials to the eigenfunctions of the so-called (quantum)
Sutherland model \cite{Su,Su2} and explain some physics terminology
which we use. We conclude this introduction with the definition of the
elliptic generalization of the Sutherland model \cite{Cal,OP} and an
outline of the rest of this paper.

\bigskip  

{\bf A definition of the Jack polynomials.} The Jack polynomials
$J_{\vn}(\vz;1/\lambda)$ are symmetric polynomials of $N>1$ variables
$\vz=(z_1,z_2,\ldots,z_N)\in\C^N$ labeled by partitions $\vn$ (i.e.\
$\vn\in\Z^N$ with $n_1\geq n_2\geq \ldots \geq n_N\geq 0$), which are
of the form
\begin{equation}
J_{\vn}(\vz) = \sum_{\vm\leq \vn} v_{\vn,\vm} M_{\vm}(\vz), \quad
M_{\vm}(\vz) = \sum_{P\in S_N}\prod_{j=1}^N z_j^{m_{Pj}} ,\label{JM}
\end{equation}
with real coefficients $v_{\vn,\vm}$ fixed up to some non-zero
normalization constant $v_{\vn,\vn}$ (which we ignore for simplicity),
and which can be defined by the property that they are eigenfunctions
of the differential operator
\begin{equation}
D= \frac1{2\lambda} \sum_{j=1}^N z_j^2 \frac{\partial^2}{\partial
z_j^2} + \sum_{\stackrel{j,k=1}{k\neq j}}^N \frac{z_j^2}{z_j-z_k}
\frac{\partial}{\partial z_j} \label{D}
\end{equation}
for some parameter $\lambda>0$; the ordering of partitions which we
use is defined as follows, $\vm\leq \vn$ if $n_1+n_2+\ldots
+n_N=m_1+m_2+\ldots m_N$ and $m_1+m_2+\ldots m_j \leq n_1+n_2+\ldots
n_j$ for all $j$, and $S_N$ is the permutation group.\footnote{Usually
these monomials $M_{\vm}$ are defined with a slightly different
normalization obtained sum summing only over the {\em distinct}
permutation $P$ \cite{St}; the choice of normalization is irrelevant
in our discussion.}  [Thus our $\lambda$, $N$, and $M_{\vn}(\vz)$
correspond to $1/\alpha$, $n$ and $m_\lambda(x)$ in \cite{St}, and we
write vectors with $N$ components in bold face.] To see that this
defines unique symmetric polynomials (up to normalization) one can
check that
\begin{equation}
D M_{\vn} = \sum_{\vm\leq \vn} b_{\vn,\vm} M_{\vm} 
\end{equation}
for certain real coefficients $b_{\vn,\vm}$ which one can compute, and
this implies that one can determine the coefficients $v_{\vn,\vm}$,
$\vm<\vn$, recursively from $v_{\vn,\vn}$ and the condition
$DJ_{\vn}=b_{\vn,\vn}J_{\vn}$; see \cite{St}, Theorem 3.1 and the
discussion thereafter. In principle one can compute straightforwardly
the Jack polynomials from this definition \cite{St}, however, the
recursion relations one thus obtains for the coefficients
$v_{\vn,\vm}$ are complicated, and it is therefore difficult to solve
them explicitly and obtain closed formulas.

\bigskip 

{\bf The Sutherland model and its relation to the Jack
polynomials.}  The Sutherland model is defined by the $N$-body
Schr\"odinger operator
\begin{equation}
H = -\sum_{j=1}^N \frac{\partial^2}{\partial x_j^2} + \gamma
\sum_{1\leq j<k\leq N} V(x_j-x_k), \label{H}
\end{equation} 
where 
\begin{equation}
V(r) = \frac1{4\sin^2(r/2)},   \label{V}
\end{equation}
$x_j\in [-\pi,\pi]$, and
\begin{equation}
\gamma=2\lambda(\lambda-1) ,\quad \lambda>0 . \label{gamma}
\end{equation}
This differential operator has a natural physical interpretation as
Hamiltonian defining a quantum mechanical model of $N$ identical
particles moving on a circle of length $2\pi$ and interacting via the
two body potentials $V$.  (To be precise, this Hamiltonian is the
self-adjoint operator on $L^2([-\pi,\pi]^N)$ defined by the Friedrichs
extension of the differential operator $H$ above, and one is only
interested in particular eigenfunctions of $H$ specified below. We set
the length of the circle to $2\pi$ only to ease notation, and an
arbitrary length $L>0$ of space as in \cite{Su2} could be easily
introduced by rescaling $x_j\to (2\pi/L) x_j$, $H\to H(2\pi/L)^2$,
etc.)  Following pioneering work by Calogero solving\footnote{By {\it
solving} a quantum many-body model we mean {\it to determine the
eigenvalues and eigenfunctions of the Hamiltonian} defining this
model.}  a similar model \cite{C}, Sutherland found an algorithm to
compute the eigenvalues and eigenfunctions of $H$ \cite{Su,Su2}, and
these exact solutions have made these models famous among theoretical
physicists (since usually such quantum models describing interacting
particles can only be studied using approximation methods; moreover,
there are many interesting physics applications and variants of these
models; see \cite{OP_review} for review).

Sutherland's solution method is closely related to our definition of
the Jack polynomials above.  To be more specific: The well-known {\em
ground state} of the Sutherland model (= eigenfunction of $H$ with the
smallest possible eigenvalue) is
\begin{equation}
\Psi_0(\vx) = \prod_{1\leq j<k\leq N} \theta(x_j-x_k)^\lambda
\label{psi0}
\end{equation}
where
\begin{equation}
\theta(r) =  \sin(r/2) , \label{theta}
\end{equation}
and to obtain the other eigenfunctions of $H$, Sutherland made the
ansatz
\begin{equation}
\Psi(\vx) = \Phi(\vx) \Psi_0(\vx)
\end{equation}
transforming the eigenvalue equation $H\Psi=\cE\Psi$ to
$H'\Phi=\cE'\Phi$ where $H'=[1/\Psi_0(\vx)]H\Psi_0(\vx)$ equals
\begin{equation}
H' = -\sum_{j=1}^N \frac{\partial^2}{\partial x_j^2} - \ii\lambda
\sum_{j<k} \frac{\ee^{\ii x_j} + \ee^{\ii x_k} }{\ee^{\ii
x_j} - \ee^{\ii x_k}}\left( 
 \frac{\partial}{\partial x_j} - \frac{\partial}{\partial x_k}
\right) , 
\label{Hp}
\end{equation}
and $\cE'=\cE-E_0$ with $E_0=\lambda^2N(N^2-1)/12$ the {\em ground
state energy} defined through $H\Psi_0=E_0 \Psi_0$.  Sutherland's
method to compute the eigenfunctions $\Phi$ of $H'$ is equivalent to
diagonalizing the operator $D$ in (\ref{D}) by making the ansatz in
(\ref{JM}) etc., as discussed above.  To see this, we change variables
to $z_j=\ee^{\ii x_j}$ and obtain
\begin{equation}
H'(\vx) = 2\lambda D(\vz) + [1-\lambda(N-1)] P(\vz) ,\quad
P=\sum_{j=1}^N z_j \frac{\partial}{\partial z_j} . 
\end{equation}
The operators $P$ and $D$ commute, and it is not difficult to see that
the eigenfunctions $\Phi_{\vn}(\vx)=J_{\vn}(\vz;1/\lambda)$ obtained
by Sutherland's method are equal to the Jack polynomials. We thus can
conclude that the eigenfunctions of the Sutherland model are given by
\begin{equation}
\Psi(\vx) = \ee^{\ii p\sum_{j=1}^N x_j}
J_{\vn}(\vz;1/\lambda)\Psi_0(\vx),\quad z_j = \ee^{\ii x_j}, 
\label{psi}
\end{equation}
with arbitrary $p\in \R$ (we used the fact that, if $\Psi(\vx)$ is an
eigenfunction of $H$, then $\exp(\ii p\sum_j x_j) \Psi(\vx)$ is an
eigenfunction of $H$ as well). The corresponding eigenvalues are found
to have the following remarkably simple formula, $\cE_0(\vn) =
\sum_{j=1}^N [ n_j + p + \lambda(N+1-2j)/2 ]^2$ \cite{Su2}. It is
interesting to note that the exponential factor on the r.h.s.\ in
(\ref{psi}) describes the center-of-mass motion of the system, which
is not very interesting and thus often ignored. However, it should be
included if one is interested in {\em all} eigenfunctions of the
model. Moreover, since transformations $p\to p+k$ and $n_j\to n_j-k$
for arbitrary integers $k\leq n_N$ leave the eigenfunction invariant,
one should restrict $p$ to integers and set $n_N=0$ if one wants to
avoid over-counting.  We also note in passing that the ansatz in
(\ref{JM}) has a natural physical interpretation: The monomials $
M_{\vn}(\vz) = \sum_{P\in S_N} \exp(\ii \sum_j n_{Pj} x_j) $ are plane
waves providing a complete set of eigenfunctions of the
non-interacting Hamiltonian $H$ with $\gamma=0$, and the Jack
polynomials thus correspond to a linear superposition of plane waves.

\bigskip

{\bf The elliptic Calogero-Sutherland model.}  In this paper
we explain a method to solve the Sutherland model which, different
from Sutherland's method, gives fully explicit formulas for the
eigenfunctions \cite{EL3,EL4}.  We will mainly concentrate on the
Sutherland model, but a main motivation for writing this paper is that
our method can be generalized to the well-known elliptic
generalization of the Sutherland model where the interaction potential
in (\ref{V}) is replaced by
\begin{equation}
V(r) = \sum_{m\in\Z} \frac1{4\sin^2[(r+\ii m\beta)/2]},\quad \beta>0
\label{eV}
\end{equation}
which is essentially the Weierstrass elliptic $\wp$-function with
periods $2\pi$ and $\ii \beta$.\footnote{To be precise:
$V(r)=\wp(r)+c_0$ where $c_0=(1/12) -(1/2)
\sum_{m\in\N}1/\sinh^2(\beta m/2)$ \cite{WW}.}  This so-called {\it
elliptic Calogero-Sutherland (eCS) model} is known to be integrable
\cite{Cal,OP}, which suggests that it should be possible to solve
it. For $N=2$ the eigenvalue equation of the eCS model is essentially
equivalent to the {\it Lam\'e equation} which was studied extensively
at the end of the 19th century; see \cite{WW} for an extensive
discussion of the classical results. The problem of finding the
general solution of the eCS model (without any restrictions on
parameters) seems to be regarded as open, even though there exist
various interesting results in this direction
\cite{DI,EK,EFK,FV1,FV2,S,T,KT,P}. As we will discuss, our method also
provides an explicit solution of the eCS model.  In the elliptic case
we will also obtain solutions as in (\ref{psi}) and (\ref{psi0})
above, only the function $\theta(r)$ is replaced by
\begin{equation}
\theta(r) = \sin(r/2) \prod_{m=1}^\infty \bigl(1-2q^{2m}\cos(r) +
q^{4m}\bigr),\quad q=\ee^{-\beta/2}, \label{etheta}
\end{equation}
which is essentially the Jacobi theta function
$\vartheta_1$,\footnote{To be precise:
$\theta(r)=\vartheta_1(r/2)/\bigl[ 2q^{1/4}
\prod_{m=1}^\infty(1-q^{2m})\bigr]$ \cite{WW}.}  and the
$J_{\vn}(\vz)$ are symmetric functions which no longer are polynomials
but reduce to the Jack polynomials in the limit $q\downarrow 0$.  We
will sketch how to obtain explicit formulas for these elliptic
generalization of the Jack polynomials by infinite
series.\footnote{These $q$-deformed Jack polynomials are different
from the Macdonald polynomials \cite{McD}.}
\bigskip

{\bf Plan of the rest of this paper.} In the next section we explain
our explicit solution of the Sutherland model and present our explicit
formulas for the Jack polynomials. Our arguments are such that they
generalize with minor changes to the elliptic case, as discussed in
Section~\ref{sec3}. This section also contains a description of our
results for the elliptic generalizations of the Jack polynomials. We
end with a few remarks in Section~\ref{sec4}.

\section{Solution of the Sutherland model}
\label{sec2}
We start by summarizing our solution in a theorem.  The proof of this
theorem has the character of a derivation which not only proves but,
as we hope, also clarifies and explains this result.

\begin{theorem} 
\label{thm1} 
{}For $\vm\in\Z^N$, let
\begin{equation} 
f_{\vm}(\vz) = \prod_{j=1}^N \left[\oint_{\cC_j} \frac{\dd \xi_j}{2\pi
\ii \xi_j} \xi_j^{m_j} \right] \frac{\prod_{1\leq j<k\leq N}
\Theta(\xi_j/\xi_k)^\lambda}{\prod_{j,k=1}^N
\Theta(z_j/\xi_k)^\lambda}
\label{fn} 
\end{equation} 
where
\begin{equation} 
\Theta(\xi) = (1-\xi) , \label{Theta}
\end{equation} 
and the integration contours $\cC_j$ are nested circles in the complex
plane enclosing the unit circle, 
\begin{equation}
\cC_j: \xi_j= \ee^{\eps j}\ee^{\ii y_j},\quad -\pi \leq y_j\leq \pi
\label{cC} 
\end{equation} 
for some $\eps>0$. Moreover, for partitions $\vn$, let
\begin{equation}
P_{\vn}(\vz) = \sum_{\vm\in\Z^N} \alpha_{\vn}(\vm) f_{\vm}(\vz)\label{Pf}
\end{equation} 
with 
\begin{eqnarray}
\alpha_{\vn}(\vm) = \delta(\vn,\vm) + \sum_{s=1}^\infty \gamma^s
\sum_{j_1<k_1} \sum_{\nu_1} \nu_1 \cdots \sum_{j_s<k_s} \sum_{\nu_s}
\nu_s \nonu \times 
\frac{\delta(\vm,\vn+\sum_{r=1}^s\nu_r\vE_{j_rk_r})}{\prod_{r=1}^s
\Bigl[ \cE_0(\vn + \sum_{\ell=1}^r \nu_\ell\vE_{j_\ell k_\ell})
-\cE_0(\vn) \Bigr]}
\label{alpha} 
\end{eqnarray} 
where
\begin{equation}
\cE_0(\vm) = \sum_{j=1}^N \Bigl[ m_j + \half\lambda(N+1-2j)\Bigr]^2
\label{cE0}
\end{equation} 
for all $\vm\in\Z^N$, $\vE_{jk}$ is the vector in $\Z^N$ with the
following components,
\begin{equation}
(\vE_{jk})_\ell = \delta_{j,\ell} -\delta_{k,\ell}  
\end{equation} 
for $j,k,\ell=1,2,\ldots,N$, and $\delta(\vn,\vm)\equiv \prod_{j=1}^N
\delta_{n_j,m_j}$. Then
\begin{equation}
\Psi_{\vn}(\vx) = P_{\vn}(\vz) \Psi_0(\vx),\quad z_j = \ee^{\ii x_j} ,  
\end{equation}
with $\Psi_0(\vx)$ defined in (\ref{psi0}) and (\ref{theta}), is an
eigenfunction of the Sutherland Hamiltonian $H$ in (\ref{H}) and
(\ref{V}), 
\begin{equation}
H\Psi_{\vn}(\vx) = \cE_{\vn} \Psi_{\vn}(\vx), \label{Hpsi}
\end{equation} 
and the corresponding eigenvalue is
\begin{equation} 
\cE_{\vn} = \cE_0(\vn).\label{EE}
\end{equation}
\end{theorem} 

It is important to note that the sums in (\ref{Pf}) and (\ref{alpha})
are finite (i.e.\ all but a finite number of terms in these sums are
zero), and the $f_\vm$ are symmetric polynomials. From this we can
conclude:

\begin{lemma} 
\label{lemma0}
The $P_{\vn}(\vz)$ given in Theorem~\ref{thm1} are symmetric
polynomials.
\end{lemma} 

{}From this one can conclude that the $P_{\vn} (\vz)$ are (essentially;
see below) equal to the Jack polynomials unless there is an
degeneracy, i.e., unless there exists a $\vm\neq \vn$ with
$\sum_j(m_j-n_j)=0$ such that $\cE_0(\vm)= \cE_0(\vn)$ \cite{EL3}. To
resolve this uncertainty due to possible degeneracies we have recently
checked that the $P_{\vn} (\vz)$ also are eigenfunctions of the
well-known third order differential operator commuting with Sutherland
Hamiltonian \cite{OP},\footnote{E.L., unpublished.} and we thus have
convinced ourselves that always,
\begin{equation}
(z_1z_2\cdots z_N)^k P_{\vn - k\ve} (\vz) = {\rm c}_{\vn,k}
J_{\vn}(\vz) , \quad \ve=(1,1,\ldots,1) \label{Jack}
\end{equation} 
with some normalization constant ${\rm c}_{\vn,k}$.  We thus get an
infinite number of explicit formulas for each Jack polynomial.  We
checked in the simplest case $N=2$ that the fact that the l.h.s.\ of
(\ref{Jack}) is independent of $k$ is non-trivial. Thus (\ref{Jack})
implies an infinite number of non-trivial identities which are
essentially the contends of Theorem~5.1 in \cite{St}.

It is interesting to note that the functions $f_\vm$ given above can
be non-zero even for certain non-partition $\vm\in\Z^N$, and in our
proof of Theorem~\ref{thm1} it does not seem essential to restrict to
$\vn$'s which are partitions. We checked for $N=2$ that the functions
$P_\vn$ all vanish unless $\vn$ is a partition, and this is true due
to highly non-trivial cancellations.\footnote{These computations were
done with the help of MATHEMATICA.} We do not know how to prove this
analytically and for all $N$ using only our approach, and the same is
true for the question of completeness, i.e., whether our construction
gives all eigenfunctions or not. However, these facts can be proven by
comparing our solution with Sutherland's \cite{Su,Su2}.

An important aspect of our method is that we expand our eigenfunctions
in a set of functions $f_\vm$ which are much more complicated than the
monomials $M_{\vm}$. However, as we will discuss, there exist fully
explicit formulas for the $f_{\vm}$.  Moreover, the $f_\vm$ are much
closer to the exact eigenfunctions than the monomials $M_{\vm}$ in the
sense that the coefficients $\alpha_{\vn}(\vm)$ are much simpler than
the coefficients $v_{\vn,\vm}$ discussed in the introduction, and we
therefore can compute their explicit series representation in
(\ref{alpha}). This later formula is our main result in this paper in
addition to our previous results in Reference \cite{EL3}.

\begin{proof}[Proof of Theorem \ref{thm1}]
We derive the result stated in Theorem~\ref{thm1} in three steps.

\bigskip

{\bf Step 1: A remarkable identity.}  The starting point of
our solution method is a particular functional identity.

\begin{lemma}\label{lemma1} 
Let
\begin{equation}
F(\vx,\vy) = \frac{\prod_{1\leq j<k\leq N} \theta(x_j-x_k)^\lambda
\theta(y_k-y_j)^\lambda}{\prod_{j,k=1}^N \theta(x_j-y_k)^\lambda}
\end{equation} 
with $\theta(r)$ the function defined in (\ref{theta}) and
$\vx,\vy\in\C^N$. Then
\begin{equation}
H(\vx)F(\vx;\vy) = H(\vy)F(\vx;\vy) , \label{Id} 
\end{equation}
where $H$ is the differential operator defined in (\ref{H}) and
(\ref{V}) acting on different arguments $\vx$ and $\vy$, as indicated.
\end{lemma} 

This can be proven by a straightforward but somewhat tedious
computation using the well-known identity
\begin{equation}
\cot(x)\cot(y) + \cot(x)\cot(z) + \cot(y)\cot(z) = 1 \quad \; \mbox{ if
$z+y+z=0$};
\end{equation} 
see Appendix A in \cite{EL3}.  While this provides an elementary
proof, it does not explain why this identity is true. We thus mention
that we found this identity by studying a particular quantum field
theory model, and this provides a natural physics interpretation of
this result; see \cite{EL_Karlstad} for review.

\begin{remark}
It is interesting to note that the identity in (\ref{Id}) can also be
obtained as a corollary of Proposition~2.1 in Reference \cite{St}.
Moreover, it seems that our method is closely related to the solution
of the Sutherland model by separation of variables \cite{KMS} since
(\ref{Id}) seems to give an alternative proof of the crucial
Theorem~4.1 in \cite{KMS}.
\end{remark}

\bigskip

{\bf Step 2: Fourier-type transformation of the remarkable
identity.} It is useful to note that the identity in (\ref{Id})
remains true if we replace $F(\vx;\vy)$ by
\begin{equation}
F'(\vx;\vy) = c \, \ee^{\ii P \sum_{j=1}^N(y_j-x_k)}F(\vx;\vy)\label{Fp}
\end{equation}
for arbitrary constants $P,c$ (this is not difficult to verify
\cite{EL3}). Below we will find it convenient to choose
\begin{equation}
P = -\lambda N/2,\quad c = (2/\ii)^{\lambda[ N(N-1)/2 - N^2] } .
\end{equation}

The idea now is to trade the variables $\vy$ for suitable quantum
numbers $\vn$ by taking the Fourier transform of (\ref{Id}) (with $F$
replaced by $F'$) with respect to the variables $\vy$, i.e., apply
$(2\pi)^{-N} \int \dd^N y \, \exp(\ii \tilde \vn\cdot \vy)$ with
suitable Fourier variables $\tilde \vn$. We need to do this with care
since, firstly, the function $F'(\vx;\vy)$ and the Hamiltonian
$H(\vy)$ have singularities and branch cuts, and secondly, the
function $F'(\vx;\vy)$ is not periodic in the variables $y_j$ but
changes by phase factors under $y_j\to y_j+2\pi$. As we will show
below in more detail, the second problem can be accounted for by
choosing the Fourier modes as
\begin{equation} 
\tilde n_j = n_j + \half\lambda (N+1-2j) ,\quad n_j\in\Z, \label{Pj}
\end{equation} 
while the first problem is solved by shifting the $y_j$-integrations
in the complex plan as follows, $y_j=\varphi_j+\ii j \eps$ with real
$\varphi_j\in [-\pi,\pi]$ and some $\eps>0$ (one can take the limit
$\eps\downarrow 0$, but this turns out to be unnecessary). Using the
fact that $1/[4\sin^2(y/2)] = -\sum_{\nu=1}^\infty \nu \ee^{\ii \nu
y}$ for $\Im(y)>0$, straightforward computations leads to the
following result.

\begin{lemma}\label{lemma2} 
{}For all $\vn\in\Z^N$, the function
\begin{equation}
\hat F(\vx;\vn) = \left[ \prod_{j=1}^N \int_{-\pi +\ii j\eps}^{\pi +
\ii j \eps} \frac{\dd y_j}{2\pi} \, \ee^{\ii \tilde n_j y_j} \right]
F'(\vx;\vy),\quad \eps>0, \label{hatF}
\end{equation} 
$\tilde n_j$ in (\ref{Pj}), and $\vx\in[-\pi,\pi]^N$, are well-defined
and obey the identities
\begin{equation} 
H(\vx)\hat F(\vx;\vn) = \cE_0(\vn)\hat F(\vx;\vn) -\gamma \sum_{j<k}
\sum_{\nu\in\Z} S_\nu \hat F(\vx;\vn + \nu \vE_{jk}) \label{HhatF}
\end{equation} 
with $\cE_0(\vn)$ and $\vE_{jk}$ defined in Theorem~\ref{thm1} and
\begin{equation}
S_0=0,\quad S_\nu = \nu \; \mbox{ and } \; S_{-\nu} = 0 \quad
\forall\nu>0.
\label{Snu} 
\end{equation}
Moreover,
\begin{equation}
\hat F(\vx;\vn) = f_{\vn}(\vz) \Psi_0(\vx) \label{hatF1}
\end{equation} 
with the functions $f_{\vn}$ and $\Psi_0$ defined in
Theorem~\ref{thm1}.

\end{lemma} 

Note that the integral in (\ref{hatF}) is independent of $\eps>0$ (due
to Cauchy's theorem).

It is not difficult to understand how (\ref{HhatF}) results from
(\ref{Id}): the l.h.s.\ is obvious since $H(\vx)$ commutes with the
Fourier transformation, while the r.h.s.\ arises from a simple
computation using
\begin{equation}
H(\vy)= -\sum_{j=1}^N \frac{\partial^2}{\partial y_j^2} -\gamma
\sum_{j<k} \sum_{\nu\in\Z} S_\nu \ee^{\ii\nu (y_j-y_k)}\quad \; \mbox{
for $\Im(y_j-y_k)>0$ }, \label{Hy}
\end{equation} 
with the first term in the r.h.s.\ of (\ref{HhatF}) coming from the
derivative term and partial integrations using $\tilde
\vn^2=\cE_0(\vn)$, and the second term comes from the interaction.

To show that $\hat F(\vx;\vn)$ is well-defined we use $ \sin(y/2)
= \half \ii \ee^{-\ii y/2}(1-\ee^{\ii y})$ to write (at this point we
make a convenient choice for the constants $c,P$ in (\ref{Fp}))
\begin{equation}
\ee^{\ii \tilde \vn\cdot \vy} F'(\vx;\vy) = \prod_{j=1}^N \xi^{n_j}
\frac{\prod_{1\leq j<k\leq N}
\Theta(\xi_j/\xi_k)^\lambda}{\prod_{j,k=1}^N\Theta(\ee^{\ii
x_j}/z_k)^\lambda} \Psi_0(\vx)
\end{equation} 
with $\Theta(\xi)=(1-\xi)$ and $\xi_j=\ee^{\ii y_j}$, where the shifts
$\lambda(N+1-2j)/2$ in the pseudo-momenta exactly cancel the terms
which would make the integrand non-analytic in the $\xi_j$. This
implies (\ref{hatF1}).

\bigskip

{\bf Step 3: Ansatz for eigenfunctions and solution of
recursion relation.}  Equation (\ref{HhatF}) suggests the following
ansatz for the eigenfunctions of the Sutherland Hamiltonian $H$,
\begin{equation} 
\Psi_{\vn}(\vx) = \sum_{\vm\in\Z^N} \alpha_{\vn}(\vm) \hat F(\vx;\vm) 
\label{ansatz}
\end{equation} 
with the normalization condition
\begin{equation}
\alpha_\vn(\vm) = \delta(\vm,\vn) + O(\gamma) . 
\label{ann}
\end{equation} 
This, (\ref{HhatF}), and the eigenvalue equation (\ref{Hpsi}) lead to
the following relations for the coefficients $\alpha_{\vn}(\vm)$,
\begin{equation}
[\cE_0(\vm) -\cE_{\vn}] \alpha_{\vn}(\vm) =
 \gamma\sum_{j<k}\sum_{\nu\in\Z} S_\nu \alpha_{\vn}(\vm - \nu\vE_{jk})
 \equiv \gamma(\opS\alpha_\vn)(\vm) \label{an}
\end{equation} 
(the last equality defines a convenient shorthand notation), i.e., the
latter relations imply (\ref{Hpsi}). We now observe that (\ref{an})
and (\ref{ann}) yield
\begin{equation}
\alpha_{\vn}(\vm) = \delta(\vm,\vn) + \gamma (R_\vn\alpha_{\vn})(\vm) 
\label{recur}
\end{equation}
with the linear operator $R_\vn$ defined as follows,
\begin{equation}
(R_\vn\alpha)(\vm) \equiv \frac1{[\![\cE_0(\vm)-\cE_{\vn}]\!]_{\vn}}
(\opS\alpha)(\vm);
\end{equation}
here and in the following we use the following convenient notation
\begin{equation}
\frac1{[\![\cE_0(\vm)-\cE_{\vn}]\!]_{\vn}} \equiv
[1-\delta(\vm,\vn)]\frac1{[\cE_0(\vm)-\cE_{\vn}]}.\label{bb}
\end{equation}
Equation (\ref{recur}) can be easily solved by iteration setting
$\alpha^{(0)}_\vn(\vm)= \delta(\vm,\vn)$,\footnote{We realized this
only when rereading our paper \cite{EL3} during this spring.} 
\begin{eqnarray}
\alpha_{\vn}(\vm) = \sum_{s=0}^\infty\gamma^s (R_\vn^s
\alpha^{(0)}_{\vn})(\vm) = \delta(\vm,\vn) + \sum_{s=1}^\infty
\gamma^s \sum_{j_1<k_1} \sum_{\nu_1\in\Z} S_{\nu_1}\cdots\nonu \times  
\sum_{j_s<k_s} \sum_{\nu_s\in\Z} S_{\nu_s} 
\frac{\delta(\vm,\vn+\sum_{r=1}^s\nu_r\vE_{j_rk_r})}{\prod_{r=1}^s
\bigl[\!\bigl[ \cE_0(\vn + \sum_{\ell=1}^r \nu_\ell\vE_{j_\ell k_\ell})
-\cE_\vn \bigr]\!\bigr]_\vn} . \label{coeff}
\end{eqnarray}
Setting $\vm=\vn$ we obtain $\alpha_\vn(\vn)=1$ (since
$(R_\vn\alpha_\vn)(\vn)=0$), and this and Equations (\ref{ann}) and
(\ref{an}) for $\vm=\vn$ imply the following condition determining the
eigenvalue $\cE_\vn$,
\begin{eqnarray}
\cE_{\vn} - \cE_0(\vn) = - \gamma (\opS \alpha_{\vn})(\vn) = -
\sum_{s=1}^\infty \gamma^{s+1} \sum_{j_1<k_1} \sum_{\nu_1\in\Z}
S_{\nu_1} \cdots \nonu\times \sum_{j_{s+1}<k_{s+1}}
\sum_{\nu_{s+1}\in\Z} S_{\nu_{s+1}}
\frac{\delta(\vzero,\sum_{r=1}^{s+1}\nu_r\vE_{j_rk_r})}{\prod_{r=1}^{s}
\bigl[\!\bigl[ \cE_0(\vn + \sum_{\ell=1}^r \nu_\ell\vE_{j_\ell
k_\ell}) -\cE_{\vn} \bigr]\!\bigr]_\vn} .  \label{cE}
\end{eqnarray}
We now recall that $S_\nu=0$ for $\nu\leq 0$, and this simplifies the
previous formula considerably: it implies that all terms in the sum on
the r.h.s.\ vanish (since the Kronecker deltas always give zero), and
we thus get the result in (\ref{EE}).  Moreover, for the same reason
we can replace all double brackets $[\![\cdots ]\!]_\vn$ in
(\ref{coeff}) by normal bracket $[\cdots]$ (since we only have the
expressions in (\ref{bb}) with $\vm\neq \vn$). Thus the series for
$\alpha_\vn(\vm)$ simplifies to the expression given in
(\ref{alpha}). \qedhere
\end{proof}

A simpler argument to derive (\ref{EE}) is as follows: since
$S_{\nu\leq 0} =0$, (\ref{an}) has an obvious triangular structure
\cite{EL3}, and inserting $\vm=\vn$ thus immediately implies
(\ref{EE}). However, this argument does not generalize to the elliptic
case, whereas ours does.

\begin{proof}[Proof of Lemma \ref{lemma0}]
The functions $f_{\vn}$ in (\ref{fn}) and (\ref{Theta}) can be
computed by Taylor expanding the integrand using the binomial series
and computing the $\xi_j$-integrals which project out the
$\xi_j$-independent terms of the integrand. One thus obtains (for
details see \cite{EL3}, Appendix B.3),
\begin{equation}
\label{p1}
f_{\vn}(\vz) = \sum_{\vm} p_{\vn,\vm} M_{\vm}(\vz)
\end{equation}
with the $M_{\vm}(\vz)$ in (\ref{JM}), and the coefficients are
\begin{equation}
\label{p2}
p_{\vn,\vm} = \sum \prod_{1\leq j'<k'\leq
N}\binom{\lambda}{\mu_{j'k'}}\prod_{j,k=1}^N \binom{-\lambda}{ \nu_{j
k}} (-1)^{\mu_{j'k'}+ \nu_{j k} } \,
\end{equation}
where the sum $\sum$ here is over all non-negative integers 
$\mu_{j'k'},\nu_{jk}$ restricted by the following $2N$ equations, 
\begin{equation}
\label{p3}
n_j = \sum_{\ell =1}^N \nu_{\ell j} + \sum_{\ell=1}^{j-1} \mu_{\ell j}
- \sum_{\ell=j+1}^{N} \mu_{j\ell} , \quad m_j = \sum_{\ell =1}^N 
\nu_{j \ell}
\end{equation}
and $m_1\geq m_2\geq \ldots \geq m_N\geq 0$. 
This shows that the $f_{\vn}$ are symmetric polynomials which are 
non-zero only if
\begin{equation}
\label{bed}
n_j+n_{j+1} + \ldots + n_N \geq 0 \quad \forall j=1,2,\ldots N ,
\end{equation} 
and the sum in (\ref{p3}) only contains terms with 
\begin{equation}
\label{p4}
\sum_{j=1}^N m_j = \sum_{j=1}^N n_j  . 
\end{equation}
The latter implies that all series in (\ref{p1}) and (\ref{alpha}) are
finite (i.e., they truncate after a finite number of terms). We
finally need to show that the denominators in (\ref{alpha}) are always
non-zero. For that we write
\begin{equation}
\vm=\vn+\vmu\;\mbox{ with } \;\vmu = \sum_{j<k} \mu_{jk} \vE_{jk}, 
\label{vmu}
\end{equation} 
and with that and (\ref{cE0}) we compute 
\begin{equation}
\cE_0(\vm)-\cE_0(\vn) = \sum_{j=1}^N \left( 2\sum_{k=j+1}^N
\mu_{jk}[n_j-n_k + (k-j)\lambda] + \Bigl[ \sum_{k<j}\mu_{kj} -
\sum_{k>j}\mu_{jk}\Bigr]^2 \right)\label{cEcE}
\end{equation} 
which is manifestly positive if $\vn$ is a partition and all
$\mu_{jk}\geq 0$ with $\vmu\neq \vzero$. Since obviously only such
terms appear in the denominators in (\ref{coeff}) we conclude that all
$\alpha_{\vn}(\vm)$ in (\ref{alpha}) are well-defined finite
series.\qedhere
\end{proof} 

As already mentioned, in our proof it does not seem essential to
restrict to $\vn$'s which are partitions. In fact, we only used this
property to have all energy differences $\cE_0(\vm)-\cE_0(\vn)$ in
(\ref{cEcE}) manifestly positive, but we only would have needed that
these energy differences are always non-zero which is true also for
certain non-partition $\vn$'s, e.g., for $N=2$ and $\lambda$
non-integer.  We feel that this point would deserve a better
understanding.

\section{Solution of the elliptic Calogero-Sutherland model: Outline}
\label{sec3}
We formulated Theorem~\ref{thm1} and its proof so that it
straightforwardly extends to the elliptic case: one only needs to
replace $V$, $\theta$, $\Theta$ and $S_\nu$ by their elliptic
generalizations. To be more specific we describe the generalizations
of steps 1--3 in our proof of Theorem~\ref{thm1} in more detail.

\bigskip

{\bf Step 1:} {\it Lemma~\ref{lemma1} holds true as it
stands with $H$ the eCS Hamiltonian defined in (\ref{H}) and
(\ref{eV}) and the function $\theta$ defined in (\ref{etheta}).}

\medskip

This can be proved by a brute-force computation using the following
well-known identify for the Weierstrass elliptic functions $\xi$ and
$\wp$ \cite{WW}, 
\begin{equation}
[\xi(x)+\xi(y)+\xi(z)]^2 =\wp(x)+\wp(y)+\wp(z)\quad \; \mbox{ if $x+y+z=0$;}
\end{equation} 
see Appendix~A in \cite{EL5}.  A quantum field theory proof of this
results can be found in \cite{EL2}.

\bigskip

{\bf Step 2:} {\it Lemma~\ref{lemma2} holds true as it
stands with $S_\nu$ in (\ref{Snu}) replaced by
\begin{equation} 
S_0=0,\quad S_\nu = \nu\frac{1}{1-q^{2\nu}} \; \mbox{ and } \; S_{-\nu} =
\nu\frac{q^{2\nu}}{1-q^{2\nu}} \quad \forall\nu>0
\label{eSnu} 
\end{equation} 
and the functions $f_{\vn}$ and $\Psi_0$ in (\ref{fn}) and
(\ref{psi0}) with
\begin{equation}
\Theta(\xi) = (1-\xi)\prod_{m=1}^\infty (1-q^{2m}\xi) (1-q^{2m}/\xi)
\label{eTheta}
\end{equation} 
and $\theta$ in (\ref{etheta}).}

\medskip

The proof is essentially unchanged, only now the functions $\Theta$
and the coefficients $S_\nu$ in the identities
\begin{equation} 
\theta(y) = \half \ii \ee^{-\ii y/2}\Theta(\ee^{\ii y})
\end{equation} 
and 
\begin{equation}
V(y) = -\sum_{\nu=1}^\infty S_\nu \ee^{\ii \nu y}\quad
\; \mbox{ for $\Im(y)>0$ }
\end{equation} 
need to be changed to what is given above, and we also need to assume
$\eps<\beta/N$; see Appendix B in \cite{EL3} for a detailed proof.
\bigskip

{\bf Step 3:} The argument given remains unchanged until and including
(\ref{cE}). The crucial difference now is that $S_\nu$ no longer
vanishes for negative values of $\nu$, and thus the above-mentioned
triangular structure is lost. Due to this (\ref{cE}) does not simplify
but remains as an implicit equation determining the eigenvalue
$\cE_{\vn}$, and thus the eigenvalues for $q>0$ are much more
complicated.

We can summarize the solution of the eCS model
thus obtained as follows.

\begin{theorem}
The eigenvalues $\cE_{\vn}$ of the eCS model are determined by
(\ref{cE}) with $S_\nu$ in (\ref{eSnu}) and $\cE_0(\vm)$ in
(\ref{cE0}). The corresponding eigenfunctions are given in
(\ref{ansatz}) with $\hat F(\vx;\vn)$ in (\ref{hatF1}),
(\ref{fn}),(\ref{eTheta}), (\ref{psi0}), and (\ref{etheta}), and the
coefficients $\alpha_{\vn}(\vm)$ in (\ref{coeff}).
\end{theorem} 

It is interesting to note that (\ref{cE}) can be turned into a fully
explicit formula for the eigenvalues as follows: Defining the function
\begin{eqnarray}
\Phh_{\vn}(\xi) \equiv - \sum_{s=0}^\infty \gamma^{s+1} \sum_{j_1<k_1}
\sum_{\nu_1\in\Z} S_{\nu_1} \cdots \sum_{j_{s+1}<k_{s+1}}
\sum_{\nu_{s+1}\in\Z} S_{\nu_{s+1}}\nonu \times
\frac{\delta(\vzero,\sum_{r=1}^{s+1}\nu_r\vE_{j_rk_r})}{\prod_{r=1}^{s}
\Bigl[\!\Bigl[ \cE_0(\vn + \sum_{\ell=1}^r \nu_\ell\vE_{j_\ell
k_\ell}) - \xi \Bigr]\!\Bigr]_\vn} \label{G1}
\end{eqnarray}
of one complex variable $\xi$, we can write (\ref{cE}) as
\begin{equation}
\cE_{\vn} = \cE_0(\vn)+\Phh_\vn(\cE_{\vn}) . \label{G2}
\end{equation} 
Using Lagrange's theorem as stated in \cite{WW}, Paragraph 7.32, the
latter equation can be solved by the following infinite
series,
\begin{equation}
\cE_{\vn} =\cE_0(\vn) + \left. \sum_{n=1}^\infty \frac1{n!}
\frac{\partial^{n-1}}{\partial \xi^{n-1}} \Phh_{\vn}(\xi)^n
\right|_{\xi=\cE_0(\vn)+a}  \label{cEseries}
\end{equation}
where $a$ is a real parameter which formally can be set to any value
such that $\Phh_\vn(\xi)$ is non-singular in $\xi=\cE_0(\vn)+a$. From
the explicit formula for $\Phh_{\vn}(\xi)$ given above it is
straightforward to compute the coefficients in this series explicitly
\cite{EL4}. In a similar manner one can find fully explicit formulas
for all coefficients $\alpha_{\vn}(\vm)$.

The arguments sketched above are enough to obtain an explicit solution
in the sense of formal power series in $q$. However, it is obviously
important to also check if the analyticity properties of the function
$\Phi_\vn(\xi)$ are such that Lagranges' theorem gives {\em
convergent} series.  For that the parameter $a$ in (\ref{cEseries}) is
important: One can prove that, if $a$ is such that
\begin{equation}
\exists \Delta>0: \forall \vm\neq \vn\; \mbox{ with } \sum_{j=1}^N
(m_j-n_j)=0:\quad |\cE_0(\vm )-\cE_0(\vn)-a|>\Delta , \label{NoR}
\end{equation}
then the function $\Phi_\vn(\xi)$ is analytic in a disc around
$\xi=\cE_0(\vn)+a$ which, in a {\em finite} $q$-interval, is large
enough for the series in (\ref{cEseries}) to converge. A similar
results holds true for all the coefficients $\alpha_\vn(\vm)$. We are
confident now to be able to prove that the eigenfunctions $\Psi_\vn$
are square integrable in a finite $q$-interval as well.

The parameter $\Delta$ is obviously important since it restricts the
range of convergence of our series solution.  It is not easy to give
general lower bounds for $\Delta$, and we only mention one simple
case: if $\lambda$ is integer then $\Delta\geq 1/2$ for $a=1/2$. It is
possible to increase $\Delta$ by finding `better' values of $a$, and
we believe that this is useful for computing $\cE_\vn$ numerically
using our results.

We plan to give a more detailed discussion of this solution of the eCS
model together with detailed proofs in a future revision of Reference
\cite{EL5}.

\section{Final remarks} 
\label{sec4}
{\bf 1.} It would be desirable to compare our perturbative solution of
the eCS model with brute-force numeric solutions and thus check if our
results are also numerically useful.\footnote{I thank David
Gomez-Ullate for stressing this point.}  

\bigskip

{\bf 2.} As discussed, the eigenvalues of the eCS model are determined
by the implicit equation in (\ref{G2}). While the explicit series
solution in (\ref{cEseries}) can be regarded as perturbative solutions
which continuously deforms the solution for $q=0$, we do not see any
reason to rule out the possibility that there are additional
non-perturbative solutions for larger values of $q$. The spectrum of
the eCS Hamiltonian as a function of $q$ might therefore be quite
complicated with qualitative changes at certain critical values of $q$
(this should be closely related to the intriguing analytical structure
of the functions $\Phh_{\vn}(\xi)$ defined above).  If so the eCS
model would challenge notions of quantum integrability based on the
simplicity of the spectrum.

\bigskip

{\bf 3.} Lemma \ref{lemma0} has an interesting
generalization.

\begin{lemma} 
{}For non-negative integers $N,M$, let 
\begin{equation}
F_{N,M}(\vx,\vy) = \frac{\prod_{1\leq j<k\leq N}
\theta(x_j-x_k)^\lambda \prod_{1\leq j<k\leq M}
\theta(y_k-y_j)^\lambda}{\prod_{j=1}^N
\prod_{k=1}^M\theta(x_j-y_k)^\lambda}
\end{equation} 
with $\theta(r)$ in (\ref{theta}) and $\vx\in\C^N$,
$\vy\in\C^M$. Then
\begin{equation}
\left[ H_{\lambda,N}(\vx) - H_{\lambda,M}(\vy) -c_{N,M} \right]
F_{N,M}(\vx;\vy) = 0 \label{Id1}
\end{equation}
where $H=H_{\lambda,N}(\vx)$ is the differential operator defined in
(\ref{H})--(\ref{gamma}) and $c_{N,M} = \lambda^2(N-M)[(N-M)^2-1]/12$.
Similarly, the function
\begin{equation}
\tilde F_{N,M}(\vx,\vy) = \prod_{1\leq j<k\leq N}
\theta(x_j-x_k)^\lambda \prod_{1\leq j<k\leq M}
\theta(y_k-y_j)^{1/\lambda} \prod_{j=1}^N \prod_{k=1}^M\theta(x_j-y_k)
\end{equation} 
obeys the identity
\begin{equation}
\left[ H_{\lambda,N}(\vx) +\lambda H_{1/\lambda,M}(\vy) -\tilde
c_{N,M} \right] \tilde F_{N,M}(\vx;\vy) = 0 \label{Id2}
\end{equation}
with $\tilde c_{N,M} = [\lambda^2N(N^2-1) + M(M^2-1)/\lambda +
3MN(\lambda N+M)]/12$.
\end{lemma} 

We thus can have different particle numbers for the two sets of
variables $\vx$ and $\vy$, and in addition to the case where the
coupling $\lambda$ is the same there is also a dual relation with
reciprocal couplings for the $\vx$ and $\vy$ variables. Note that, for
$M=0$, (\ref{Id1}) and (\ref{Id2}) both reduce to the eigenvalue
equation $H\Psi_0=E_0\Psi_0$ for the groundstate in (\ref{psi0}).

We found these identities using a quantum field theory construction,
but once known they can also be proven by brute-force computations
\cite{EL6}. Taking the Fourier transform of (\ref{Id1}) and
(\ref{Id2}) etc.\ (as in Section~\ref{sec2}), one can obtain
additional explicit formulas, and thus identities, for the Jack
polynomials. We suspect that (\ref{Id2}) would lead to an alternative
proof of Theorem~3.3 (duality relation of the Jack polynomials) in
\cite{St}.

It is important to note that the generalizations of (\ref{Id1}) and
(\ref{Id2}) to the elliptic case exist but involve, in general, a term
with a $\beta$-derivative \cite{EL6}, and this term is only absent in
the special case stated in Lemma~\ref{lemma1}.  We only know how to
construct eigenfunctions when this $\beta$-derivative term is absent,
which is the reason why we only discussed this case in detail.

\bigskip

{\bf 4.} The method explained in this paper can be adapted to give
explicit solutions of the original Calogero model \cite{C} and the
$BC_N$-variants of the Calogero-Sutherland models
\cite{OP_review}.\footnote{M.\ Halln\"as and E.\ Langmann, work in
progress.}  It will be interesting to see if this method can also
provide an alternative solution of the Ruijsenaars models \cite{R} and
thus can give interesting explicit formulas for Macdonald's
generalization of the Jack polynomials \cite{McD}; see e.g.\
\cite{AOS} for interesting results in this direction.

\section*{Acknowledgements}
This paper was written during a scientific gathering at the {\it
``Centro Internacional de Ciencias A.C.''} (CIC) in Cuernavaca
(Mexico) devoted to Integrable Systems. I would like to thank the
organizers Francesco Calogero and Antonio Degasperis for inviting me
to this meeting and the CIC for hospitality. I am grateful to Vadim
Kuznetsov and Evgueni Sklyanin for helpful discussions and for
reviving my interest in the eCS model. I thank Martin Halln\"as for
helpful comments on the manuscript.

\bibliographystyle{amsalpha} 


\providecommand{\bysame}{\leavevmode\hbox to3em{\hrulefill}\thinspace}
\providecommand{\href}[2]{#2}

\end{document}